\documentclass[aps,twocolumn]{revtex4}
\usepackage{epsfig}
\newcommand{\be}{\begin{equation}}
\newcommand{\ee}{\end{equation}}
\newcommand{\ba}{\begin{array}}
\newcommand{\ea}{\end{array}}

\begin{document}
\title{Scaling of Star Polymers with one to 80 Arms}
\author{Hsiao-Ping Hsu, Walter Nadler, and Peter Grassberger}
\affiliation{John-von-Neumann Institute for Computing, Forschungszentrum
J\"ulich, D-52425 J\"ulich, Germany}

\date{\today}
\begin{abstract}
We present large statistics simulations of 3-dimensional star polymers with 
up to $f=80$ arms, and with up to 4000 monomers per arm for small values of 
$f$. They were done for the Domb-Joyce model 
on the simple cubic lattice. This is a model with soft core exclusion which
allows multiple occupancy of sites but punishes each same-site pair 
of monomers with a Boltzmann factor $v<1$. We use this to allow all
arms to be attached at the central site, and we use the `magic' value $v=0.6$
to minimize corrections to scaling. The simulations are made with a very 
efficient chain growth algorithm with resampling, PERM, modified to allow 
simultaneous growth of all arms. This allows us to measure not only the swelling 
(as observed from the center-to-end distances), but also the partition sum. 
The latter gives very precise estimates of the critical exponents $\gamma_f$.
For completeness we made also extensive simulations of linear (unbranched)
polymers which give the best estimates for the exponent $\gamma$.

\end{abstract}
\maketitle

\section{Introduction}

Star polymers are of interest both for their technical applications, ranging 
from lubricant additives to paints \cite{applic,applic2}, and for the theoretical 
challenge which they represent. Polymer theory in general is one of the prime 
fields where renormalization group theory can be used and compared in detail 
with real experiments \cite{degennes,cloizeaux}. The simplest non-trivial 
objects in this respect are the partition sum and the r.m.s. end-to-end 
distance of a single long flexible linear (unbranched) polymer with $N$ monomers in a good solvent, 
which scale as 
\be
   Z_N \sim \mu^{-N} N^{\gamma-1}     \label{ZN}
\ee
and 
\be
   R_N^2 \approx A_1 N^{2\nu}\;.      \label{RN}
\ee
Star polymers, i.e. $f$ such chains linked together at a single point, are 
some of the simplest examples of polymers with non-trivial topology. As 
shown by Duplantier \cite{dupla}, all such polymer networks are characterized 
by equations similar to Eqs.~(\ref{ZN}) and (\ref{RN}), with the critical fugacity 
$\mu$ and the critical exponent $\nu$ being the same for all topologies,
but with $\gamma$ being universal only within each topology. For star 
polymers composed of $f$ arms of length $N$ each, one has in particular
\be
   Z_{N,f} \sim \mu^{-fN} N^{\gamma_f-1} \label{ZNF}
\ee
and 
\be
   R_{N,f}^2 \approx A_f N^{2\nu}\;,   \label{RNF}
\ee
where $R_{N,f}$ is the r.m.s. Euclidean center-to-end distance.

The behaviour of $\gamma_f$ and of the {\it swelling factor} $A_f$ are of 
central interest, both for finite $f$ and for $f\to\infty$. In two dimensions,
$\gamma_f$ can be calculated exactly using conformal invariance \cite{dupla},
but no exact results are known for $d=3$. Renormalization group methods
give $\epsilon$ expansions up to third power in $\epsilon = 4-d$
\cite{schaefer-ferber-dupl}, but these 
are non-convergent power series and have to be resummed before being 
applicable in $d=3$. The results are debated, in particular for large values
of $f$ \cite{ferber-holov}. For the swelling factor the situation is similarly 
unclear. Phenomenologists tend to compare with predictions based on Gaussian
(i.e. free) chains \cite{zimm-stock} or on heuristic assumptions 
\cite{daoud-cotton,birshtein}. There exist several renormalization group 
calculations, but those not based heavily on simulation data 
\cite{miyake-freed,douglas-freed} seem to describe some of the data rather 
poorly, and Monte Carlo simulations are needed to fix 
free parameters in such theories \cite{lue-kiselev1,lue-kiselev2}.

In view of this, Monte Carlo 
\cite{barrett,batoulis,lue-kiselev1,zifferer,shida,cecca-freire,ohno}
and molecular dynamics \cite{grest-kremer-witten,grest} simulations have played a 
major role in the efforts to understand the behaviour of star polymers. 
Molecular dynamics simulations \cite{grest} have indeed been used to study 
very large stars, with up to 80 arms of length $N=100$ each, but it is not 
clear whether these simulations have really reached equilibrium. Moreover, 
both molecular dynamics and Monte Carlo methods with fixed chain lengths
(including the pivot algorithm \cite{lue-kiselev1,zifferer}) cannot measure
the partition sum and thus give no information on $\gamma_f$. For the latter 
one has to use chain growth methods \cite{barrett,batoulis,zifferer,shida,ohno,footnote}.
Unfortunately, with the methods used so far it has not been possible to go 
beyond 24 arms \cite{shida}, and even these were too short and the data were
too noisy to provide a clear cut picture of the asymptotic behaviour.

We decided therefore to perform simulations with several improvements
which allow us to reach much larger systems and much higher accuracy. 
To obtain a good estimate for $\mu$ and for the critical exponents of 
unbranched polymers, we also made extensive simulations of linear chains.
The model and the method of simulation are described in the next section. 
Results are given in Sec.~3, while we end with a discussion in Sec.~4.

\section{Model and Method}

Let us first describe in detail our model. For efficiency, and since we are 
only interested in scaling behaviour, we use a lattice model. Indeed, we use 
the simplest version, the simple cubic lattice. But instead of simulating self 
avoiding walks as in previous works, we simulate the Domb-Joyce model \cite{dj}
at its `magic' interaction strength $v=v^*$. In the Domb-Joyce model polymers
are described by lattice walks where monomers sit at sites and are connected
by bonds of length 1. Multiple visits to the same site are allowed, but for 
any pair of monomers occupying the same site one has a repulsive energy $\epsilon>0$
giving rise to a Boltzmann factor $v=\exp(-\beta\epsilon) < 1$. The partition 
sum of a linear chain molecule of $N+1$ monomers is thus a sum over all walks of 
$N$ steps, each weighted with $v^m$ where $m$ is the total number of pairs 
occupying the same site, $m = \sum_{i<j}\delta_{{\bf x}_i,{\bf x}_j}$. For star
polymers we studied in the present work two variants. In both variants arms of 
$N$ monomers are attached to a central site. In the first variant, the central 
site is singly occupied. In the second, it is occupied by $f$ monomers, one for 
each arm. We studied both variants in order to verify that results were independent
of this detail, and we include in our final error estimates the uncertainty it 
entails.

Using the Domb-Joyce model has two main advantages. First of all, it allows us
to attach a large number of arms to a point-like center. In the present work, 
we go up to $f=80$~\cite{footnote1}.
Previously, authors had used extended cores. Although these 
cores were much smaller than the radii of the polymers themselves and should thus
not destroy the asymptotic scaling, they do introduce a finite length scale and 
present therefore corrections to scaling terms which complicate the analysis.

More important is that there is one special (`magic') value of $v$, called $v^*$
in the following, where corrections to scaling are minimal and where asymptotic
scaling laws are reached fastest. For single chains it has been estimated 
\cite{gss, belohorec} as $v^* \approx 0.6$ with rather small error bars, and we 
shall in the following assume this value to be exact. In the renormalization 
group language, the flow of the effective Hamiltonian to its fixed point 
in the stable manifold of the latter contains one direction of slowest approach.
For a generic starting point there is a non-zero component in this direction,
which then determines the leading correction to scaling. If one starts however 
with the flow such that this component is absent, the approach to scaling is 
governed by the next-to-leading correction term and is much faster. A similar 
observation has been made also for off-lattice bead-rod models with fixed 
bond length, where the leading corrections to scaling are absent for a certain
`magic' ratio between bead size and rod length \cite{baumg,krueger,lue-kiselev0}.

Since the value of $v^*$ should depend only on the internal structure of the 
chains, for star polymers it should be independent of the number of arms. For 
the bead-rod model this was carefully verified in \cite{lue-kiselev1}. We thus 
simulated only with $v=v^*$.

For our simulations we used the pruned-enriched Rosenbluth method (PERM)
\cite{g97}. This is a biased chain growth algorithm, similar to the 
Rosenbluth-Rosenbluth \cite{rr55} method. In the latter, the bias induced
by avoiding double occupancy is compensated by a weight factor which is basically 
of entropic origin. In the present case, we have both a bias compensating 
factor and a Boltzmann factor, the product of which tends to fluctuate
wildly if there is no perfect importance sampling. These fluctuations are 
suppressed in PERM by pruning low weight configurations and cloning those 
with high weight. Indeed, any bias can be employed in PERM, as long as it 
can be compensated by a weight factor. In previous simulations of diluted
polymers we use a Markov approximation called {\it Markovian anticipation}
\cite{cylinder,strip,slab}. In the present case we did not expect this to be 
very useful, because the main interactions are not within one arm but between 
different arms. Thus we used instead a very simple bias where each arm tends
to grow preferentially outward (except for the simulations for $f=1$ where we 
used of course Markovian anticipation). The strength of this bias was adjusted 
by trial and error. It decreased with the length of the arm and increased with 
$f$. Details will not be given since they are not very important, and working 
without this bias would have increased the errors by only a factor 
$\approx 2$, in general.

A final comment is that it is easy to modify the basic PERM algorithm given e.g.
in the appendix of \cite{g97} such that all $f$ arms are grown simultaneously
\cite{coluzzi}. This is done by having $f$ growth sites ${\bf x}_1,\ldots {\bf x}_f$.
Chain growth is made in PERM by calling recursively a subroutine for each monomer 
addition.  For multi-arm growth, we add an integer $k\in [1,\ldots f]$ to the 
argument list of this subroutine, such that a subroutine called itself with 
argument $k$ calls the next subroutine with $(k\; {\rm mod}\; f)+1$. In this way a 
monomer is added to each arm before the next round of monomers is added. Compared 
to a scheme where one arm is grown entirely before the next arm is started, the 
main advantage is that each chain grows in the field of all the others, and is 
thus, by the population control (pruning/cloning), guided  to grow into the 
correct outward direction. If chains were grown one after the other, this bias 
would be absent for the first chains which then would grow into ``wrong" directions,
resulting in very low weight configurations.

\section{Results}

\subsection{Partition Sums and $\gamma$-Exponents}

One of the outstanding features of chain growth methods such as PERM is that they
give estimates for the partition sum. Indeed, these estimates are a basic part
of the simulations, since the population control is based on these estimates.

According to Eq.~(3) we expect $Z_{N,f} \mu^{fN}$ to approach a power law 
$const \; N^{\gamma_f-1}$ at large $N$. One precise way to estimate $\gamma_f$ 
is to subtract a term $a_f \ln N$ from $\ln (Z_{N,f}\mu^{fN})$, and adjust the 
constant $a_f$ such that the difference gives a flat curve for large $N$, when 
plotted against $\ln N$. This gives then $\gamma_f = 1+a_f$. Alternatively, we 
could plot $\ln Z_{N,f} - \ln Z_{fN,1} - a_f' \ln N$ against $\ln N$, in which 
case a flat curve is obtained when $a' = \gamma_f - \gamma$. We prefer both
methods over a least square fit, say, since they allow directly to check visually
for the presence of corrections to scaling. If such corrections seem needed, 
one can subtract them and obtain in this way the most reliable and precise
estimates of $\gamma_f$ -- remembering of course that estimating a critical 
index involves an extrapolation and is thus ill-posed, giving at best subjective
error estimates.

For either method we need precise estimates of the partition sum of linear
chains. We thus performed first extensive simulations of linear ($f=1$) 
Domb-Joyce chains, creating altogether $\approx 4\times 10^8$ chains of length
$N=8000$. In Fig.~1 we plot effective exponents obtained from triple ratios 
\cite{gss} $Z_{aN}^x Z_{bN}^y/Z_N$. Here $a$ and $b$ are chosen such as to 
minimize statistical and systematic errors \cite{gss}, and powers $x$ and $y$
are fixed such that $\mu$ and the overall normalization drop out. With $a=1/3$
and $b=5$ we have 
\be
   \gamma_{\rm eff}(N) = 1 +{7\ln Z_N - 6 \ln Z_{N/3} - \ln Z_{5N}\over \ln (3^6/5)}
\ee
which is plotted against $1/N^{0.96}$. The fact that we find essentially a straight 
line (apart from odd/even oscillations due to the special structure of the cubic lattice) 
indicates that the leading correction to scaling exponent is $\Delta \approx 0.96$ 
which is much larger than the value $\Delta \approx 1/2$ for generic self avoiding 
walks, indicating that $v = 0.6$ is indeed close to the magic value. Our 
estimate is therefore
\be
   \gamma = \lim_{N\to\infty} \gamma_{\rm eff}(N) = 1.1573\pm 0.0002 \;.
\ee
This is in good agreement with the best previous estimates \cite{causo,gss}
but more precise. Using this estimate we obtain then 
\be
   \mu = 0.18812145\pm 0.00000003.
\ee

\begin{figure}
  \begin{center}
   \psfig{file=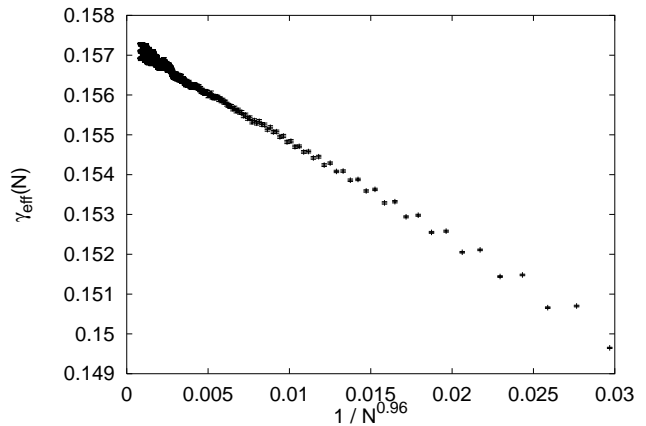,width=5.9cm,angle=270}
   \caption{Effective exponents $\gamma_{\rm eff}(N)$ for linear ($f=1$) `magical'
    Domb-Joyce polymers against $1/N^{0.96}$.}
\label{gamma_1}
\end{center}
\end{figure}

\begin{figure}
  \begin{center}
   \psfig{file=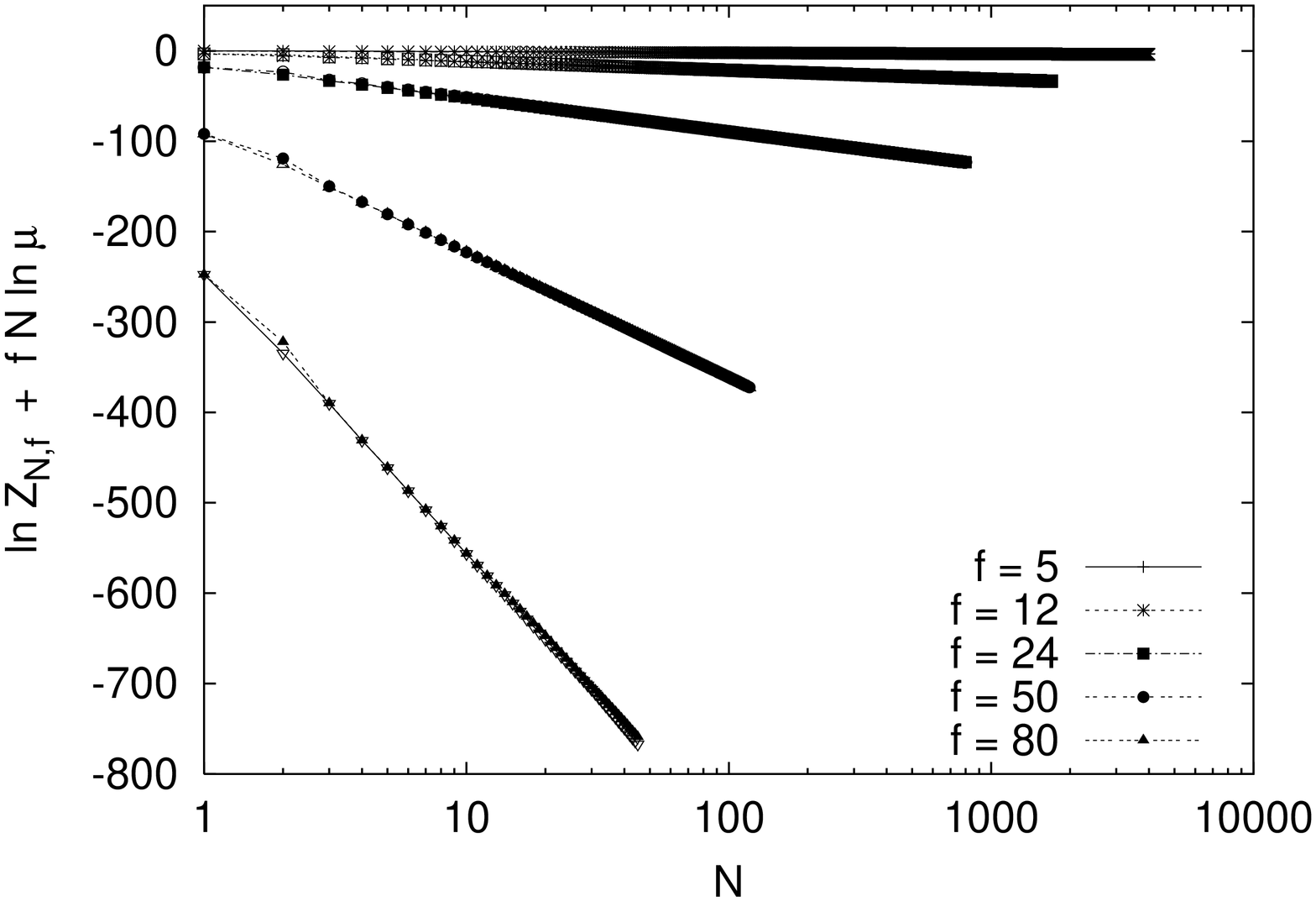,width=5.9cm,angle=270}
   \caption{Logarithms of the partition functions $Z_{N,f}$ multiplied by 
    $\mu^{fN}$. For each pair of close-by lines, the upper refers to singly
    occupied centers, while the lower one has $f$ monomers located at the 
    center.}
\label{Zgamma_1}
\end{center}
\end{figure}

\begin{figure}
  \begin{center}
   \psfig{file=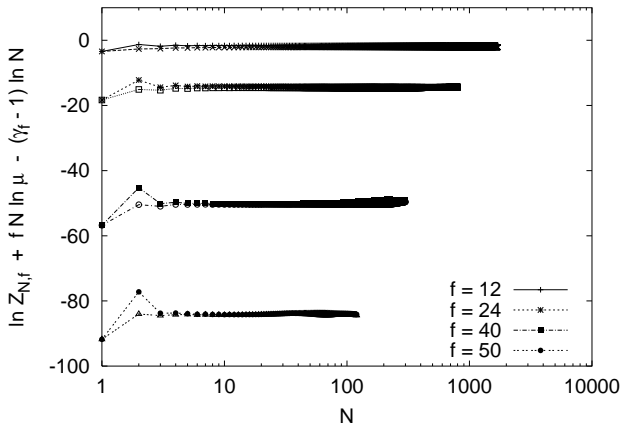,width=5.9cm,angle=270}
   \caption{Logarithms of the partition functions $Z_{N,f}$ multiplied by 
    $\mu^{fN}$, plus $(1-\gamma_f) \ln N$. For each pair of close-by lines, 
    the upper refers again to singly occupied centers, while the lower one 
    has $f$ monomers located at the center.}
\label{ZZgamma_1}
\end{center}
\end{figure}

After having obtained a precise estimate for $\mu$, we can now discuss the 
results for stars. Results for a few selected values of $f$ are shown in 
Fig.~2. We plot there $\ln Z_{N,f} +fN \ln \mu$ for both variants, i.e. the
center singly occupied or $f$ times occupied. The latter gives smaller values 
of $Z_{N,f}$, but the difference is visible only for $N=2$. For larger $N$ 
both agree, except for $f=80$ and large values of $N$ where our sampling 
algorithm starts to break down.

For a precise estimate of $\gamma_f$ we of course did not use plots like Fig.~2,
but we subtracted $(\gamma_f-1)\ln N$ as explained above. Then we see (Fig.~3)
that there are non-negligible corrections to scaling, but our arms are long 
enough so that our estimates of $\gamma_f$ are not affected by them. Our final
results, obtained by averaging over both variants of the model, are shown in 
Table~1 and in Fig.~4. In Table~1 we also give additional information such as 
the arm lengths and the total statistics. We also list previous estimates for 
comparison. We see reasonable agreement in general, although those previous estimates 
which were quoted with error bars \cite{batoulis} are off by many standard 
deviations. We should add that the simulations in \cite{batoulis} involved 
much shorter chains and lower statistics.

\begin{figure}
  \begin{center}
   \psfig{file=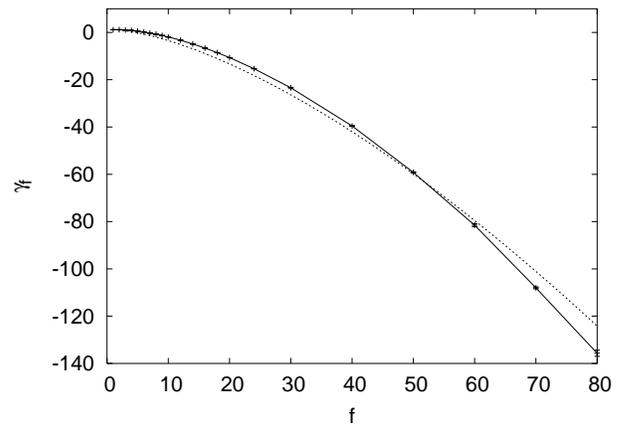,width=5.9cm,angle=270}
   \caption{Exponents $\gamma_f$ versus $f$. The full line is just a polygon
    connecting the points, the dashed line is a fit with the large-$f$ 
    behaviour as predicted by the cone approximation, Eq.~(\ref{cone}).  }
\label{fig-gamma_f}
\end{center}
\end{figure}

\begin{table*}
\begin{center}
\caption{Main results. } 
  \label{table1}
\begin{ruledtabular}
\begin{tabular*}{7.cm}{@{\extracolsep\fill}|r|r|r|r@{.}l|c|r@{.}l|c|}
  $f$& $N\;$ &       runs$\;\;$   &\multicolumn{2}{c|}{$\gamma_f$}& previous     &\multicolumn{2}{c|}{$A_f / A_1$}& previous     \\
     &       &                    &\multicolumn{2}{c|}{}          & estimates    &\multicolumn{2}{c|}{}           & estimates    \\ \hline
   1 &$\;$8000 &  $486\times 10^6$  &   1&\hspace{-0.87cm}1573(2)  & 1.1575(6)$^a$     &    1&\hspace{-0.88cm}0       &              \\
   2 &  4000 &   $71\times 10^6$  &   1&\hspace{-0.87cm}1573(2)  &                     &    1&\hspace{-0.88cm}0614(5) & $1.0628^f$  \\
   3 &  4000 &   $83\times 10^6$  &   1&\hspace{-0.87cm}0426(7)  & 1.089(1)$^b$        &    1&\hspace{-0.88cm}1123(5) & $1.1139^f; 1.128^g$  \\
   4 &  4000 &  $142\times 10^6$  &   0&\hspace{-0.87cm}8355(10) & 0.879(1)$^b$        &    1&\hspace{-0.88cm}1553(6) & $1.1581^f$  \\
   5 &  4000 &  $114\times 10^6$  &   0&\hspace{-0.87cm}5440(12) & 0.567(2)$^b$        &    1&\hspace{-0.88cm}1939(8) &              \\
   6 &  3000 &   $73\times 10^6$  &   0&\hspace{-0.87cm}1801(20) & $0.16(1)^b;0.14^c$  &    1&\hspace{-0.88cm}2295(9) & $1.2322^f; 1.265^g$  \\
   7 &  2500 &   $73\times 10^6$  &  -0&\hspace{-0.87cm}2520(25) & $-0.33,-0.20^c$     &    1&\hspace{-0.88cm}2626(11)&              \\
   8 &  2300 &   $59\times 10^6$  &  -0&\hspace{-0.87cm}748(3)   & $-0.88,-0.60^c;-1.00^d$ &    1&\hspace{-0.88cm}2934(12)& $1.2951^f$  \\
   9 &  2150 &   $48\times 10^6$  &  -1&\hspace{-0.87cm}306(5)   & $-1.51,-1.01^c$     &    1&\hspace{-0.88cm}3225(14)&              \\
  10 &  2000 &   $67\times 10^6$  &  -1&\hspace{-0.87cm}922(7)   &                     &    1&\hspace{-0.88cm}3494(16)& $1.3519^f; 1.424^g$  \\
  12 &  1700 &   $73\times 10^6$  &  -3&\hspace{-0.87cm}296(9)   & -3.35$^d; -3.4(3)^e$  &    1&\hspace{-0.88cm}4014(17)& $1.4017^f$  \\
  14 &  1400 &   $66\times 10^6$  &  -4&\hspace{-0.87cm}874(9)   & -4.94$^d$           &    1&\hspace{-0.88cm}4481(19)&              \\
  16 &  1200 &   $96\times 10^6$  &  -6&\hspace{-0.87cm}640(10)  & -5.90$^d$           &    1&\hspace{-0.88cm}4917(24)&              \\
  18 &  1100 &   $96\times 10^6$  &  -8&\hspace{-0.87cm}575(12)  & -8.12$^d; -8.9(2)^e$  &    1&\hspace{-0.88cm}532(3)  &              \\
  20 &  1000 &  $130\times 10^6$  & -10&\hspace{-0.87cm}66(2)    & -11.33$^d$          &    1&\hspace{-0.88cm}574(4)  & $1.660^g$    \\
  24 &   800 &  $147\times 10^6$  & -15&\hspace{-0.87cm}32(4)    & -18.13$^d$          &    1&\hspace{-0.88cm}643(5)  &              \\
  30 &   500 &  $316\times 10^6$  & -23&\hspace{-0.87cm}40(6)    &                     &    1&\hspace{-0.88cm}735(7)  & $1.896^g$    \\
  40 &   300 &  $880\times 10^6$  & -39&\hspace{-0.87cm}55(13)   &                     &    1&\hspace{-0.88cm}883(14)   & $2.036^g$    \\
  50 &   120 &  $1194\times 10^6$ & -59&\hspace{-0.87cm}2(2)     &                     &    1&\hspace{-0.88cm}95(2)   & $2.208^g$    \\
  60 &    80 &  $1712\times 10^6$ & -81&\hspace{-0.87cm}5(4)     &                     &    2&\hspace{-0.88cm}04(3)   &              \\
  70 &    61 &  $1944\times 10^6$ &-108&\hspace{-0.87cm}0(7)     &                     &    2&\hspace{-0.88cm}13(4)   &              \\
  80 &    45 &$\;1966\times 10^6$ &$\;$-135&\hspace{-0.87cm}7(13)&                     &    2&\hspace{-0.88cm}16(6)   &              \\ 
\end{tabular*}
$^a$ Ref.\cite{causo}, Monte Carlo\\
$^b$ Ref.\cite{batoulis}, Monte Carlo\\
$^c$ Ref.\cite{schaefer-ferber-dupl}, $\epsilon$-expansion\\
$^d$ Ref.\cite{shida}, Monte Carlo\\
$^e$ Ref.\cite{ohno}, Monte Carlo\\
$^f$ Ref.\cite{zifferer}, Monte Carlo (tetrahedral lattice)\\
$^g$ Ref.\cite{grest}, Molecular dynamics (off-lattice; values for $N=50$)
\end{ruledtabular}
\end{center}
\end{table*}

Previous theoretical predictions of $\gamma_f$ used $\epsilon = 4-d$ - expansions
\cite{schaefer-ferber-dupl,ferber-holov} and the cone approximation 
\cite{witten,ohno-binder}. The latter assumes that each branch is confined to a cone 
of space angle $4\pi/f$, and gives 
\be 
   \gamma_f - 1 \sim - f^{3/2}\;.       \label{cone}
\ee
As seen from Fig.~4 this is not too far off, but it definitely does not provide 
a quantitative fit to our data. The best fit with a power law $\gamma_f - 1 \sim - 
(f-1.5)^z$ would be obtained with $z\approx 1.68$, but we do not claim that this
exponent has any deeper significance.

In contrast to the cone approximation which is basically heuristic and cannot be
improved systematically, $\epsilon$ - expansions have a firm theoretical basis.
But the expansion itself is at best asymptotic, and each term gives a contribution
to $\gamma_f$ which is polynomial in $f$. Thus it cannot be used without 
re-summation. Such re-summations have not yet been attempted for $f\to\infty$.
For small $f$, results are given in \cite{schaefer-ferber-dupl,ferber-holov}, 
and are listed in Table~1. They are in the right order of magnitude, but their
precision is not sufficient to draw any firm conclusion beyond the fact that 
the resummed $\epsilon$ - expansion is obviously not in conflict with the 
Monte Carlo data.

\subsection{Coil Sizes}

We measured only rms. center-to-end distances of the arms (resp. end-to-end 
distances for $f=1$). This was done `on the fly', i.e. we did not store each
configuration and measure its properties in a second step of analysis. The 
reason is that an off-line analysis would have required very large files, 
and reading a configuration from disk or tape would have been not much faster 
than creating it from scratch. We neither measured shape parameters nor radii of 
gyration, since any such additional measurement would have slowed down the 
analysis considerably, and since the main purpose of the present work was to 
demonstrate the efficiency of PERM and to study the main universal properties
of large stars.

\begin{figure}
  \begin{center}
   \psfig{file=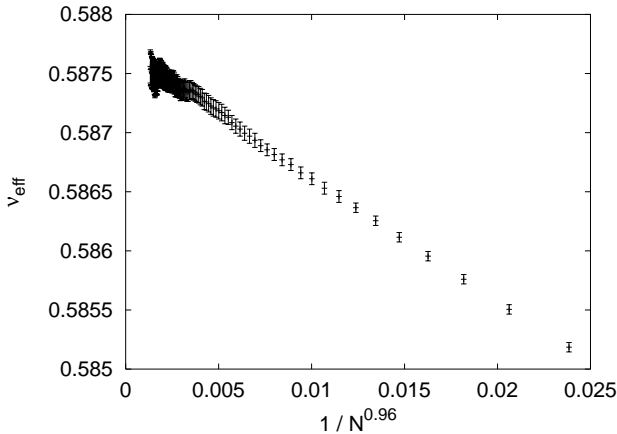,width=5.9cm,angle=270}
   \caption{Effective exponents $\nu_{\rm eff}(N)$ for linear `magical'
    Domb-Joyce polymers against $1/N^{0.96}$.}
\label{fig-nu_f}
\end{center}
\end{figure}

\begin{figure}
  \begin{center}
   \psfig{file=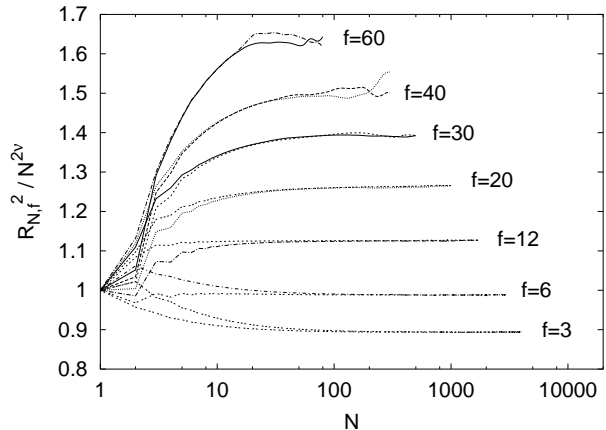,width=5.9cm,angle=270}
   \caption{$R^2_{N,f}/N^{2\nu}$ plotted versus $\ln N$, for seven selected 
    values of $f$. For each $f$, results are shown for both variants of the 
    model (single occupancy of the center: lower curve; $f$-fold occupancy: 
    upper curve). The structures seen for $f\geq 40$ and large $N$ are 
    statistical fluctuations.}
\label{fig-R_Nf}
\end{center}
\end{figure}

As for $\gamma_f$, we first need a careful analysis of linear chains, to 
obtain precise estimates of $\nu$ and of the amplitude $A_1$. In Fig.~5 we 
plot effective exponents $\nu_{\rm eff}(N) = (\ln 16)^{-1} \ln [R^2_{8N}/R^2_N]$,
again versus $1/N^{0.96}$. We see again
a straight line (as in Fig.~1), verifying again that the correction to scaling
exponent is close to 1. Extrapolating to $N\to\infty$ we find $\nu = 0.58767(20)$.
Together with previous estimates reviewed in \cite{slab}, this gives our best
estimate 
\be
   \nu = 0.58765 \pm 0.00020 \;.
\ee
Notice that this is more precise than the field theoretic estimates
obtained from the $\epsilon$-expansion~\cite{cloizeaux}.
The resulting amplitude is then 
\be 
   A_1 = \lim_{N\to\infty} R^2_N / N^{2\nu} = 0.8038 \pm 0.0005\;.
\ee

To obtain the amplitudes $A_f$ for stars we assume the value of $\nu$ as given 
above. We can then plot either $R^2_{N,f}/N^{2\nu}$ versus $N$ (which gives $A_f$ 
directly), or $R^2_{N,f}/R^2_N$ versus $N$, which gives $A_f/A_1$. To check
for systematic corrections we did both. Some typical curves obtained with the 
first method are shown in Fig.~6. For each value of $f$ we see two curves, one 
for each variant of the model: The upper curve is always that with the center
$f$ times occupied, the lower one corresponds to a singly occupied center.
For large values of $f$ ($f\geq 40$) we see large fluctuations, indicating the 
limit where our sampling breaks down. Otherwise we see large corrections to 
scaling, but they all are dominantly $\sim 1/N$, i.e. analytic corrections, 
and they have rather small influences on our final estimates of $A_f$. 

\begin{figure}
  \begin{center}
   \psfig{file=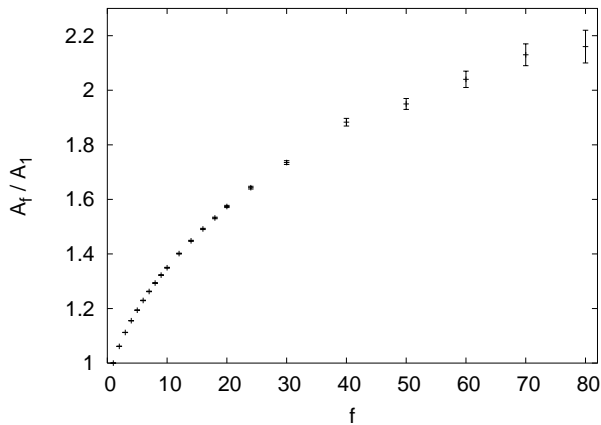,width=5.9cm,angle=270}
   \caption{Amplitude ratios $A_f/A_1$ plotted against $f$.}
      \label{fig-A_f.ps}
   \end{center}
\end{figure}

These estimates are given in Table~1 and plotted in Fig.~7. We show indeed the 
ratios $A_f/A_1$, to facilitate the comparison with previous estimates. The best 
previous estimates are those of Zifferer \cite{zifferer} and are also given in 
Table~1. We see very good agreement, even if most of the values of \cite{zifferer} 
are outside our error bars. The data of Zifferer were obtained from simulations 
on the tetrahedral lattice, and they indicate that the ratios $A_f/A_1$ are indeed 
universal. In \cite{grest}, stars with up to 80 arms were simulated off-lattice
by means of molecular dynamics. But it seems that the stars with $N=100$ were not 
equilibrated, at least for $f=1$ (see Table~1 of \cite{grest}). Therefore we 
list for comparison only the data for $N=50$. They are systematically larger than
our results and those of \cite{zifferer}, and the
discrepancy increases with $f$. This suggests that even these simulations 
had not reached equilibrium.

The most cited predictions for $A_f/A_1$ are from a heuristic blob model
\cite{daoud-cotton,birshtein}. It gives 
\be
   {R^2_{N,f}\over R^2_N}  \sim  f^{1-\nu} \sim f^{0.41}
\ee
which is in gross violation with our data. The fact that this Daoud-Cotton
model gives a too strong swelling with $f$ is well known \cite{zifferer,havrank}.
Our data cannot be fitted by a pure
power law, but asymptotically (for $f\to\infty$) they tend roughly to 
$A_f/A_1 \approx 0.78\; f^{0.235}$. Again we do not expect this to be the true asymptotic
behaviour, but it provides at least a useful guide for extrapolations. 

Renormalization group (RG) calculations of star polymer sizes have been performed
in \cite{miyake-freed,douglas-freed}, but it was already pointed out in 
\cite{lue-kiselev1,lue-kiselev2} that these have difficulties in describing the 
large-$f$ behaviour. Using their own simulations to fix some of the parameters 
in an improved RG calculation, Lue and Kiselev \cite{lue-kiselev1,lue-kiselev2} were able to 
fix these problems in the sense that their RG calculation described perfectly 
the behaviour of the penetration function \cite{lue-kiselev1,lue-kiselev2}.
Unfortunately, they did not give predictions for $A_f$, so we cannot make 
a detailed comparison with our data. But we should point out that 
\cite{lue-kiselev1,lue-kiselev2} also obtained much less swelling
with $f$ than predicted in Refs.\cite{daoud-cotton,miyake-freed,douglas-freed}.

\section{Discussion}

We have demonstrated that chains growth methods with resampling, and the 
PERM algorithm in particular, are able to produce very precise Monte Carlo 
data for star polymers with many arms. Using the Domb-Joyce model on the 
simple cubic lattice, we combined this with absence of leading corrections 
to scaling and with the possibility to connect arbitrarily many arms to a pointlike 
core. This allowed us to test conjectured scaling laws for the entropic critical
exponents $\gamma_f$ and for the $f$-dependent swelling of single arms.
In principle we could have measured during these simulations also other 
observables like monomer densities, star shapes, radii of gyration, etc.

Our most interesting results are for the exponents $\gamma_f$. All previous 
simulations were compatible with the predictions from the heuristic Daoud-Cotton 
model, but they were not very precise. There are also no good experimental 
results for these exponents, although they are fundamental for the entropy
(and thus also for the free energy) of star polymers in good solvents. Our
results show that these predictions are qualitatively correct ($\gamma_f$
is negative and diverges as $-f^\alpha$, but the exponent $\alpha$ clearly 
disagrees with the prediction. 

We also disagree with the prediction of the Daoud-Cotton model for the sizes
of star polymers, and indeed the disagreement for the end-to-center distances
is larger than for $\gamma_f$. They increase with $f$ much slower than predicted.
But this finding is not entirely new, it had been observed previously in 
Monte Carlo simulations \cite{zifferer,havrank}. Our data are compatible with
these, but more precise and extending to larger values of $f$. Disagreement with
the Daoud-Cotton prediction for star polymer sizes was also found in some 
experiments \cite{held}, but there are also repeated claims in the literature 
\cite{grest,willner} that experiments are compatible with it. We have no good 
explanation for the latter, except that the interpretation of experiments for 
diluted solutions might be less easy than anticipated. 

With slight modifications of the algorithm one can also study related 
problems like stars center-absorbed to surfaces \cite{shida2}, stars 
confined between two planar walls \cite{sikorski}, heterostars
\cite{ferber-holov,havrank}, interactions between two star polymers
\cite{witten,jusufi,rubio,lue-kiselev2}, or star polymer - colloid interactions 
\cite{jusufi2}. We expect that PERM will be more efficient than previous
algorithms (not the least because it gives immediately precise entropy 
estimates), in particular if applied to lattice models. PERM can also be 
applied off-lattice \cite{g97,still}, but its advantage is in general 
less pronounced there. We hope to present simulations for some of these 
problems in the near future.


\begin{thebibliography}{30}
\bibitem{footnote} In principle one could also use a modified pivot algorithm
   for measuring $\gamma$ exponents \cite{causo}, but this has not yet
   been done for star polymers.
\bibitem{footnote1} 
This implies that the density in the core region can
become larger than one, which might seem unphysical. But the
Domb-Joyce model is a coarse-grained model where the lattice
constant can be understood to be much larger than the
real monomer size. Densities larger than one do not present problems.
\end{thebibliography}
\end{document}